\documentclass[preprintnumbers,article,amsmath,amssymb,floatfix,10pt,prd,superscriptaddress,nofootinbib]{revtex4}
\pdfoutput=1

\usepackage{doi}%<------------
\usepackage{hyperref}
\hypersetup{
  colorlinks=true,        % false: boxed links; true: colored links
  linkcolor=blue,         % color of internal links
  citecolor=cyan,         % color of links to bibliography
}

\usepackage[pdftex]{graphicx}% Include figure files
\usepackage{epstopdf} % Allow PDF graphics with latex
\usepackage{dcolumn}% Align table columns on decimal point
\usepackage{bm}% bold math
\usepackage{color}
\usepackage{enumitem}

\newcommand{\beq}{\begin{equation}}
\newcommand{\eeq}{\end{equation}}
\newcommand{\bea}{\begin{eqnarray}}
\newcommand{\eea}{\end{eqnarray}}

\usepackage{bbm}
\usepackage{amsfonts}
\usepackage{mathrsfs}
\usepackage{latexsym}
\usepackage{epsfig}
\usepackage{epstopdf}
\usepackage{dcolumn}
\usepackage{bm}
\usepackage{color}
\usepackage{comment}
\usepackage{xcolor}
\usepackage{orcidlink}

\begin{document}
\title{Onset of Chaos and Orbital Instabilities around Rotating Weyl Black Holes}

\author{Shubham Kala\orcidlink{0000-0003-2379-0204}}
\email{shubhamkala871@gmail.com}
\affiliation{Centre for Space Research, North-West University, Potchefstroom 2520,
South Africa}
\affiliation{Research Center of Astrophysics and Cosmology, Khazar University, Baku, AZ1096, 41 Mehseti Street, Azerbaijan.}

\author{M. Yousaf\orcidlink{0000-0001-7280-177X}}
\email{myousaf.math@gmail.com}
\affiliation{Department of Mathematics, Virtual University of Pakistan, 54-Lawrence Road, Lahore 54000, Pakistan.}

\author{Pradeep Singh}
\email{pradeep.dawan@gmail.com }
\affiliation{Department of Applied Sciences, Faculty of Engineering and Technology, Gurukula Kangri (Deemed to be University), Haridwar--249404, India}

\author{Faisal Javed}
\email{faisaljaved.math@gmail.com }
\affiliation{School of Mathematical and Statistics, Zaozhuang University, Zaozhuang 277160, China}
\affiliation{Department of Mathematics, School of Science, University of Management and Technology,  Lahore, 54000, Pakistan.}

\author{Hemwati Nandan}
\email{hnandan@associates.iucaa.in}
\affiliation{Department of Physics, Hemwati Nandan Bahuguna Garhwal University,
Srinagar--246174, Uttarakhand, India}
\affiliation{Centre for Space Research, North-West University, Potchefstroom 2520,
South Africa}

\author{G. Mustafa\orcidlink{0000-0003-1409-2009}}
\email{gmustafa3828@gmail.com }
\affiliation{Department of Physics, Zhejiang Normal University,
Jinhua 321004, China}

\affiliation{Jadara University Research Center, Jadara University, Irbid21110, Jordan}

\begin{abstract}
We investigate nonlinear photon motion in the vicinity of a rotating black hole (BH) within Weyl conformal gravity (WCG) by employing Poincaré sections, Lyapunov exponents, Fast Lyapunov Indicators (FLI), Kolmogorov-Sinai (KS) entropy, and Discrete Indicator of Chaos (DIG) maps. The joint influence of BH spin and higher order curvature corrections, governed by the parameters $\xi$ and $\gamma$, induces a transition from nearly integrable trajectories to markedly chaotic dynamics in the strong field domain. As $\xi$ and $\gamma$ increase, the invariant tori are increasingly deformed and eventually break up. Resonance islands and stochastic layers then form, producing large chaotic regions in phase space, especially near the extremal spin regime. Increasing Lyapunov exponents, faster FLI growth, and higher KS entropy provide quantitative measures of the enhanced dynamical instability, the increased sensitivity to initial conditions, and the corresponding rate of information production. DIG maps show shrinking regular regions and expanding chaotic domains around the photon sphere, indicating that conformal gravity corrections destabilize null circular orbits. Taken together, these findings clarify how relativistic frame dragging and higher order curvature terms jointly control photon chaos and point to nonlinear photon dynamics as a potential probe of departures from General Relativity in the strong-gravity regime.
\end{abstract}

\maketitle

\date{\today}

%\tableofcontents

%%%%%%%%%%%%%%%%%%%%%%%%%%%%%%%%%%%%%%%%%%%%%%%%%
\section{Introduction}\label{S1}
%%%%%%%%%%%%%%%%%%%%%%%%%%%%%%%%%%%%%%%%%%%%%%%%%

The BHs are among the most fascinating phenomena and the most compact entities in the universe, and in General Relativity (GR), they are exact solutions to the equations governing the gravitational field of a distribution of mass. They occupy a region of spacetime from which nothing and no signal (even light) can escape once it crosses the event horizon. The BHs not only provide critical insights into gravitational phenomena but also serve as a natural laboratory for the study of spacetime structure, singularities, and a fundamental issue in quantum gravity. Investigations of intense gravitational fields surrounding them, particularly through electromagnetic observations, are widely discussed in the literature~\cite{bambi2017black}, whereas owing to earlier limitations in observational capabilities, they could not initially be identified directly and their existence relied primarily on indirect signatures. In 1974, Stephen Hawking demonstrated that BHs are not completely black but emit radiation, which came to be known as Hawking radiation. This radiation is a consequence of the formation of virtual particle-antiparticle pairs near the event horizon, where one particle of the pair can escape to infinity while the other is drawn into them. This process results in BH losing a small amount of mass, and eventually, it will evaporate. In the early 1970s, with Cygnus $X-1$, which was recognized as a strong candidate for a stellar mass BH~\cite{bolton1972identification}, while near infrared flares from accreting gas around the supermassive BH at the galactic centre were investigated by Genzel and his collaborators in their work \cite{genzel2003near}. Since then, numerous compact X-ray sources associated with supermassive BHs have been detected in nuclei of galaxies, and additional confirmation came from observation of gravitational waves generated during the merger of two BHs. Independent support for their existence can be found in~\cite{abbott2018erratum}. A major milestone occurred when the Event Horizon Telescope (EHT), operating as a global very long baseline interferometry array, released the first direct image of a supermassive BH located at the center of the elliptical galaxy M87~\cite{akiyama2019first,event2019first}. First Sagittarius A* EHT results and shadow of supermassive BH in center of the Milky Way studied in~\cite{akiyama2022first}, while Kala and his collaborators \cite{Kala:2025xnb} studied null geodesics, thermodynamics, weak gravitational lensing, and BH shadow characteristics of a Frolov regular BH with constraints from EHT observations.

The EHT observations capture the shadow produced by a BH against the luminous emission from its surrounding accretion flow. The dark central region corresponds to the BH shadow, while the surrounding bright ring like structure arises from photons orbiting near the unstable circular trajectories, commonly referred to as the photon ring~\cite{cunha2018shadows}. The analysis of particle and photon geodesics in the vicinity of BHs plays a crucial role in revealing both the nature of these compact objects and the geometrical structure of the surrounding spacetime. The first investigation of a BH shadow carried out by Synge~\cite{synge1966escape}, while the interaction between radiation and accreting matter such as ionized gas surrounding material significantly affects the observed optical features of the system. In 1979, Luminet \cite{luminet1979image} extended the analysis by considering a thin accretion disk and produced the first theoretical image of a BH embedded in a realistic environment by extending Synge's work. Subsequently, Bardeen investigated photon trajectories in the rotating Kerr spacetime and demonstrated that rotation distorts the shadow~\cite{bardeen1973black}. Since then, in an influential study, Gralla et al.~\cite{gralla2019black} analyzed an optically and geometrically thin accretion flow around a Schwarzschild BH and investigated the formation of bright rings surrounding the shadow. Later, Narayan and his collaborators~\cite{narayan2019shadow} examined BH images in spherical accretion environments and demonstrated that the resulting observational signatures differ markedly from thin disk configurations, where the inner disk edge leaves a prominent imprint on the observed image. Photon rings of spherically symmetric BHs and robust tests of non-Kerr metrics studied in~\cite{wielgus2021photon} and supermassive BH spin constraints from polarimetry in an equatorial disk model examined in \cite{palumbo2025supermassive} as well as lensing and shadow of a BH surrounded by a heavy accretion disk in \cite{cunha2020lensing}.

Beyond Einstein's standard GR, BH shadows have also been studied extensively within various alternative theories of gravity, both with and without accretion effects. These include shadows and photon spheres with spherical accretions in the four-dimensional Gauss-Bonnet BH~\cite{zeng2020shadows}, null geodesics as well as shadow of hairy BHs in Einstein-Maxwell-dilaton gravity prtesented in~\cite{heydari2022null}, shadows of Einstein-dilaton-Gauss-Bonnet BHs investigated in~\cite{cunha2017shadows}. Singh et al. \cite{Singh2026ChaosKerrBR} studied strong-gravity chaos in rotating Kerr-Bertotti-Robinson BHs and shadows of rotating five-dimensional charged EMCS BHs~\cite{amir2018shadows}, while the shadow and photon sphere of the charged BH in Rastall gravity~\cite{guo2021shadow}. Observational appearances of a {$f (R)$} global monopole BH illuminated by various accretions investigated in \cite{li2021observational}, including modified gravity BHs. Kala et al. \cite{Kala:2022uog} investigated shadow and weak gravitational lensing of a rotating regular BH in a non-minimally coupled Einstein-Yang-Mills theory in the presence of plasma, while Einstein aether solutions, quintessence BHs, Kerr Newman Kasuya, Kerr modified gravity and different BH solutions~\cite{Kala:2020prt,Kala:2021ppi,wang2022optical,sun2022shadows,yousaf2024fuzzy,gao2023investigating,Roy:2025qmx,Kukreti:2025rzn}. In addition, the optical appearance of BHs in regular (singularity free) spacetimes along with different related studies, broadened this line of study across a range of gravitational models attracted considerable interest in recent years~\cite{guo2022influence,asad2024evolution,Vishvakarma:2024icz,yousaf2025implications,donmez2026accretion,donmez2026relativistic,yousaf2026dual}.

Although Einstein's GR has been remarkably successful over the past century in describing gravitational phenomena ranging from planetary motion to cosmological dynamics, several conceptual and observational challenges remain. In particular, GR is expected to break down at very small (quantum) scales where a consistent quantum theory of gravity is required, and it does not possess invariance under conformal (local scale) transformations. Furthermore, explaining the nearly flat rotation curves of galaxies and the observed late-time acceleration of the universe within the GR scenario typically requires the introduction of dark matter and dark energy components~\cite{corbelli2000extended}, while improved constraints on dark energy from Chandra X-ray observations of the largest relaxed galaxy clusters studied in \cite{allen2008improved}. These limitations motivate the exploration of alternative gravitational theories that preserve the empirical successes of GR while addressing its shortcomings. Among the proposed extensions, conformal (Weyl) gravity represents a compelling candidate. WCG~\cite{weyl1918reine} represents a well-known extension of GR, while in Einstiein's GR, the field equations remain invariant under general coordinate and Lorentz transformations. Conformal gravity, however, is constructed with an additional symmetry, as invariance under local conformal (scale) transformations of the spacetime metric. comprehensive solution to the cosmological constant, zero-point energy, and quantum gravity problems examined in~\cite{mannheim2011comprehensive}.  In this theoretical scenario, the geometry of spacetime is preserved under local rescaling of lengths, implying that only angle-preserving (shape-preserving) properties of spacetime are physically meaningful~\cite{Turner:2020gxo,Varieschi:2009vlp}. This additional symmetry significantly constrains the gravitational dynamics and leads to a unique higher-order gravitational theory based on the Weyl curvature tensor, while the action of conformal gravity is built from the square of the Weyl tensor, which corresponds to the traceless part of the Riemann curvature tensor obtained after removing all contractions associated with the Ricci tensor and scalar curvature~\cite{Alestas:2019wtw}. As a consequence, the resulting field equations differ from Einstein's equations and involve higher-order derivatives of the metric. These equations can be written in a form analogous to the Einstein equations, with the Bach tensor playing the role of the geometrical gravitational tensor and the energy momentum tensor acting as the matter source~\cite{Mannheim:2005bfa}, so the theory is scale invariant and characterized by a dimensionless gravitational coupling constant.

Static and spherically symmetric BH solutions in this framework were first constructed in Buchdahl's work~\cite{buchdahl1953set}, exact vacuum solution to conformal Weyl gravity, and galactic rotation curves in~\cite{mannheim1989exact}. Heydari-Fard and his collaborators \cite{heydarifard2025observational} analyzed the shadow and optical characteristics of BHs within conformal Weyl gravity as well as they calculated radii of the event horizon, photon sphere, and shadow. Also, it was found to be larger than those of the Schwarzschild solution in GR, with the Weyl parameter controlling this enhancement. Observational data from EHT for M87* are used to constrain the free parameter of the theory, to model realistic astrophysical conditions, the BH assumed to be surrounded by an optically and geometrically thin accretion disk. By considering different inner emission radii, the contributions of direct emission, lensing rings, and photon rings to the observed image are examined. Subsequently, a wide variety of configurations obtained, including rotating and charged rotating BHs, topological solutions, static cylindrical geometries, Lifshitz BHs, and wormhole spacetimes within Weyl gravity~\cite{mannheim1991solutions,klemm1998topological,said2012exact,lu2012ads,lobo2008general,heydari2023time}. In recent years, numerous physical properties of Weyl BHs investigated, such as gravitational lensing~\cite{cattani2013correct}, BH evaporation in conformal Weyl gravity in~\cite{xu2019black}, and quasinormal modes of BHs in Weyl gravity: electromagnetic and gravitational perturbations studied in \cite{momennia2020quasinormal} as well as Stability and quasinormal modes of BHs in conformal Weyl gravity~\cite{momennia2021stability}. The cosmological consequences of conformal Weyl gravity have also been widely explored for representative studies (see for further details.~\cite{mannheim2012cosmological,nesbet2022weyl}). 

A significant feature of conformal gravity appears in its static, spherically symmetric vacuum solution obtained in~\cite{Alestas:2019wtw}, which can be regarded as the conformal gravity counterpart of the Schwarzschild spacetime. Unlike the Schwarzschild solution, the metric contains additional linear and quadratic radial terms, which introduce new parameters that are conceived naturally from the fourth order structure of the theory and lead to modifications in the gravitational potential at galactic and cosmological scales. Furthermore, rotating BH solutions were also constructed within Weyl gravity using appropriate generalization procedures~\cite{Asuncion:2025cxw}. These solutions extend the Kerr geometry by incorporating additional parameters associated with conformal symmetry. The presence of these parameters modifies the near horizon structure and photon motion around the compact object, thereby affecting observable phenomena such as accretion disk emission, gravitational lensing, and BH shadow properties. 
{Chaos and nonlinear dynamics have emerged as important tools for investigating complex gravitational and Hamiltonian systems. Chaotic behavior has been studied in a broad range of contexts, including black hole dynamics, relativistic Hamiltonian systems, and gravitational few body problems. For example, recent studies have investigated chaotic orbits in unstable fractional Hamiltonian systems and relativistic kicked Hamiltonians with modified dispersion relations, demonstrating how nonlinearities, bifurcations, and modified dynamical terms can generate complex phase-space structures and chaotic trajectories~\cite{Dodelson:2025rng,Polchinski:2015cea,el2025chaotic,el2026extended,el2026stability,El-Nabulsi:2026pls}. Furthermore, several recent contributions focusing on chaotic dynamics, Lyapunov exponents, BH thermodynamics and from the perspective of modified gravity, the analysis of Lyapunov exponents in $f(R)$ gravity BHs provided further insights into the interplay between gravitational modifications, dynamical chaos and probing strong-gravity chaos and stability of circular photon orbits around BH~\cite{Deich2024,li2025chaotic,singh2025circular,sang2026chaotic,singh2026probing,gao2026interplay,hou2026quasinormal,cheng2026lyapunov,li2026chaotic}.}

{Despite there being a significant amount of literature on chaotic photon dynamics in several BH spacetimes of GR as well as in some modified gravity theories, there is no study devoted to a detailed non-linear dynamical analysis of photon motion around BHs in WCG. Specifically, the combined effect of BH rotation and the presence of higher-order terms in the expansion of the conformal curvature tensor on the transformation of the regular photon behavior into the chaotic one has never been numerically studied with the help of complementary chaos indicators. To fill this gap, in this paper, we conduct a systematic investigation using the tools of Poincaré surfaces of section, Lyapunov exponents, Fast Lyapunov Indicators (FLI), Kolmogorov-Sinai entropy, and Discrete Indicator of Chaos (DIC) maps.}

The paper is organized as follows: In section~\ref{S2}, we present a rotating Kerr BH solution in Weyl gravity and outline the mathematical scenario for null geodesics, while section~\ref{S3} is devoted to the analysis of null circular orbits and their Lyapunov stability in Weyl gravity. In section~\ref{S4}, we construct Poincaré sections to visualize the phase space structure and characterize regular and chaotic trajectories, and section~\ref{S5} introduces the main Lyapunov stability indicators, including the Largest Lyapunov Exponent (LLE) as well as the FLI. In Section~\ref{S6}, we discuss the KS entropy as a global measure of chaos, whereas Section~\ref {S7} focuses on the quantitative detection of chaotic behavior using Weighted Birkhoff Averages. Finally, section~\ref{S8} summarizes our main results and discussion.

%%%%%%%%%%%%%%%%%%%%%%%%%%%%%%%%%%%%%%%%%%%%%%%%%%%
\section{Rotating Kerr BH in Weyl gravity}\label{S2}
%%%%%%%%%%%%%%%%%%%%%%%%%%%%%%%%%%%%%%%%%%%%%%%%%%%

As WCG \cite{Weyl:1918ib} is an alternative to the GR, the Einstein field equations are invariant under general coordinate transformations, $g_{ij}(x) \rightarrow g'_{ij}(x')$, Lorentz transformations $x^i \rightarrow \Lambda^i_{\;j} x^j$, and WCG is modified to be invariant under local conformal rescaling of the metric
\begin{equation}
g_{ij}(x) \rightarrow \tilde{g}_{ij}(x) = \Omega^2(x)\, g_{ij}(x),
\end{equation}
where $\Omega(x)$ denotes the conformal factor specifying the local scaling of the line element \cite{Turner:2020gxo, Varieschi:2009vlp}. The requirement of local conformal invariance for the gravitational field equations uniquely leads, at the level of fourth-order derivative theories, to an action functional of the form
\begin{equation}\label{eq1}
A_{\mathrm{Weyl}} = -\alpha_g \int \mathrm{d}^4 x\, \sqrt{-g}\, \mathcal{C}_{\lambda i j \kappa} \mathcal{C}^{\lambda i j \kappa},
\end{equation}
where $\alpha_{\mathrm{g}}$ is a dimensionless gravitational coupling constant, $g$ is the determinant of the metric tensor $g_{ij}$, and $\mathcal{C}_{\lambda i j \kappa}$ is the Weyl tensor. The latter is defined as the completely traceless part of the Riemann curvature tensor $R_{\lambda i j \kappa}$, obtained by subtracting all trace contributions constructed from the Ricci tensor and the Ricci scalar \cite{Alestas:2019wtw}.

\begin{eqnarray}\label{eq2}
\mathcal{C}_{\lambda i j \kappa}=\mathcal{R}_{\lambda i j \kappa}- \frac{1}{2}\left(g_{\lambda j} \mathcal{R}_{i \kappa}-g_{\lambda \kappa} \mathcal{R}_{i j}-g_{i j} \mathcal{R}_{\lambda \kappa}+g_{i \kappa} \mathcal{r}_{\lambda j}\right)+\frac{1}{6} \mathcal{R}_\alpha^\alpha\left(g_{\lambda j} g_{i \kappa}-g_{\lambda \kappa} g_{i j}\right),
\end{eqnarray}
where, as usual, $\mathcal{R}_{i j}=g^{\rho \sigma} \mathcal{R}_{\sigma i \rho j}$ denotes the Ricci tensor. By varying the gravitational action $A_{\mathrm{Weyl}}$ from Eq. (\ref{eq1}) with respect to the metric $g_{i j}$, one obtains \cite{Mannheim:2005bfa}
\begin{equation}\label{eq3}
\frac{1}{\sqrt{-g}} \frac{\delta A_{\mathrm{Weyl}}}{\delta g_{i j}}=-2 \alpha_g \mathcal{W}^{i j},
\end{equation}
where $\mathcal{W}^{i j}$ denotes the traceless Bach tensor. Equivalently, this can be expressed as
\begin{equation}\label{eq4}
\mathcal{W}_{i j}=\mathcal{W}_{i j}^{(b)}-\frac{1}{3} \mathcal{W}_{i j}^{(a)},
\end{equation}
with
\begin{eqnarray*}
\left\{\begin{array}{l}
\mathcal{W}_{i j}^{(a)}=2 g_{i j} R_{; \lambda}^{; \lambda}-2 R_{; i ; j}-2 R R_{i j}+\frac{1}{2} R^2, \\
\mathcal{W}_{i j}^{(b)}=\frac{1}{2} g_{i j} R_{; \lambda}^{; \lambda}+R_{i j}^{; \lambda}{ }_{; \lambda}-R_i^\lambda{ }_{; i ; \lambda}-2 R_{i \lambda} R_j{ }^\lambda+\frac{1}{2} g_{i j} R_{\rho \lambda} R^{\rho \lambda}.
\end{array}\right.
\end{eqnarray*}
This, in turn, permits the conformal gravity field equations to be cast in a representation that is formally analogous to the corresponding field equations of GR
\begin{equation}\label{eq5}
\mathcal{W}_{i j}=\frac{1}{4 \alpha_{\mathrm{g}}} T_{i j}.
\end{equation}
{Several fundamental deviations of conformal gravity from GR become apparent upon analysis of the static, uncharged, spherically symmetric vacuum solution obtained by Mannheim and Kazanas \cite{mannheim1989exact}}. We refer to this spacetime metric, which may be regarded as the conformal gravity analogue of the Schwarzschild solution, as
\begin{equation}\label{eq6}
\mathrm{d} s^2=-D(r) \mathrm{d} t^2+\frac{\mathrm{d} r^2}{D(r)}+r^2\left(\mathrm{~d} \theta^2+\sin ^2 \theta \mathrm{~d} \phi^2\right).
\end{equation}
By introducing the dimensionless mass parameter \(\xi = GM / c^{2}\) and the associated mass parameter
\begin{equation}\label{eq7}
\widetilde{m} \equiv \xi\left(1-\frac{3}{2} \xi \gamma\right).
\end{equation}
We represent the lapse function in the form
\begin{equation}\label{eq8}
D(r)=1-3 \xi \gamma-\frac{2 \widetilde{m}}{r}+\gamma r-\kappa r^2.
\end{equation}
The constants $\gamma$ and $\kappa$ arise as supplementary parameters as a direct consequence of the fourth-order nature of conformal gravity. By employing the procedure described in Ref.~\cite{Asuncion:2025cxw}, one can derive the rotating Kerr BH solution in Weyl gravity in the following form:
\begin{eqnarray}\label{mainbh}
\mathrm{d} s^2 &=&-\left(1-\frac{2 \widetilde{m} r}{\rho^2}-k\left(r^2-a^2 \cos ^2 \theta\right)\right) \mathrm{d} t^2 +2\left(\frac{-2 \widetilde{m} r a \sin ^2 \theta+k a\left(a^2\left(r^2+a^2\right) \cos ^4 \theta-r^4 \sin ^2 \theta\right)}{\rho^2}\right) \mathrm{d} t \mathrm{~d} \phi \nonumber \\
&+&\left(\frac{\rho^2}{\Delta^{\mathrm{H}}}\right) \mathrm{d} r^2+\left(\frac{\rho^2}{\Delta^\theta}\right) \mathrm{d} \theta^2+\left(\frac{\Sigma^2 \sin ^2 \theta}{\rho^2}\right) \mathrm{d} \phi^2,
\end{eqnarray}
with the following auxiliary terms, which are defined as:
\begin{eqnarray*}
 k &\equiv & \kappa+\frac{\gamma^2(1-\xi \gamma)}{(2-3 \xi \gamma)^2},\;\;\;
 \rho^2 \equiv  r^2+a^2 \cos ^2 \theta,\;\;\;
 \Delta^{\mathrm{H}} \equiv -k r^4+r^2-2 \widetilde{m} r+a^2,\;\;\;
 \Delta^\theta \equiv  1-k a^2 \cos ^2 \theta \cot ^2 \theta \\
\Sigma^2 &\equiv &\Delta^\theta\left(r^2+a^2\right)^2-a^2 \Delta^{\mathrm{H}} \sin ^2 \theta.
\end{eqnarray*}

{The motivation for adopting metric, (i.e., defined in Eq.~(7) corresponds to the well-known static, spherically symmetric vacuum solution in conformal Weyl gravity originally obtained by Mannheim and Kazanas \cite{mannheim1989exact}) lies in its ability to incorporate additional linear and quadratic curvature terms, characterized by the parameters $\gamma$ and $\kappa$, which encode deviations from GR. These terms play a crucial role in modifying the gravitational potential at both galactic and strong field scales, furthermore, this metric serves as the foundational solution from which the rotating Weyl BH spacetime (used in Eq.~(10)) is constructed. Therefore, its inclusion is essential for establishing the theoretical scenario of our analysis and for investigating the imprint of conformal gravity corrections on photon dynamics and chaotic behavior. A detailed analysis of the horizon structure and photon region for the spacetime considered here has already been presented in the more recent work~\cite{zulqarnain2026probing}. There, the parameter space was systematically explored to identify physically admissible regimes, ensuring the existence of well-defined horizons and photon regions.}

%%%%%__________________________________SECTION_____________________________________________%%%%
\subsection{Mathematical framework for null geodesics in Weyl gravity}
%%%%%______________________________________________________________________________________%%%%
{The study of the mathematical framework for null geodesics is essential for understanding photon dynamics in the vicinity of BHs, in particular, null geodesics govern the propagation of light and therefore determine key observational features such as photon spheres, BH shadows, and gravitational lensing. In the context of our considered gravity (i.e., Weyl gravity), analyzing null geodesics allows us to investigate how conformal corrections modify the effective gravitational potential and influence the stability of photon orbits. The formulation of null geodesics provides the fundamental dynamical equations required to compute Lyapunov exponents, construct Poincar\'e sections, and evaluate other chaos diagnostics.} 
The rotating BH solution in WCG is described by the line element given in Eq.~(\ref{mainbh}). The nonvanishing components of the metric tensor are
\begin{align}
g_{tt} &= -\left(1-\frac{2\widetilde{m} r}{\rho^2}-k(r^2-a^2\cos^2\theta)\right),\\
g_{t\phi} &= \frac{-2\widetilde{m}ra\sin^2\theta+k a\!\left[a^2(r^2+a^2)\cos^4\theta-r^4\sin^2\theta\right]}{\rho^2},\\
g_{rr} &= \frac{\rho^2}{\Delta^{\rm H}}, \qquad 
g_{\theta\theta}=\frac{\rho^2}{\Delta^\theta},\\
g_{\phi\phi} &= \frac{\Sigma^2\sin^2\theta}{\rho^2},
\end{align}
where the functions \(\rho^2\), \(\Delta^{\rm H}\), \(\Delta^\theta\), and \(\Sigma^2\) are defined in Eq.~(\ref{mainbh}).
The geodesic motion is governed by the Lagrangian
\begin{equation}
\mathcal{L}=\frac12 g_{\mu\nu}\dot{x}^\mu\dot{x}^\nu ,
\end{equation}
with conjugate momenta
\begin{equation}
p_\mu=g_{\mu\nu}\dot{x}^\nu .
\end{equation}
Because the spacetime is stationary and axisymmetric, the coordinates \(t\) and \(\phi\) are cyclic, giving rise to two constants of motion,
\begin{equation}
p_t=-E, \qquad p_\phi=L,
\end{equation}
where \(E\) and \(L\) denote the conserved energy and angular momentum of the photon.
Solving for \(\dot t\) and \(\dot\phi\), we obtain
\begin{align}
\dot t &= \frac{g_{\phi\phi}E + g_{t\phi}L}{g_{t\phi}^2-g_{tt}g_{\phi\phi}},\\
\dot\phi &= \frac{-g_{t\phi}E - g_{tt}L}{g_{t\phi}^2-g_{tt}g_{\phi\phi}}.
\end{align}
Using the identity
\begin{equation}
g_{t\phi}^2-g_{tt}g_{\phi\phi}=\Delta^{\rm H}\sin^2\theta ,
\end{equation}
the above expressions simplify to
\begin{align}
\dot t &= \frac{g_{\phi\phi}E + g_{t\phi}L}{\Delta^{\rm H}\sin^2\theta},\\
\dot\phi &= \frac{-g_{t\phi}E - g_{tt}L}{\Delta^{\rm H}\sin^2\theta}.
\end{align}
Restricting the motion to the equatorial plane (\(\theta=\pi/2\)), the null condition
\begin{equation}
g_{\mu\nu}\dot{x}^\mu\dot{x}^\nu=0,
\end{equation}
yields the radial equation of motion
\begin{equation}
\dot r^2=\frac{1}{r^4}\left[(E(r^2+a^2)-aL)^2-\Delta^{\rm H}(L-aE)^2\right],
\label{rdot_Weyl}
\end{equation}
where
\begin{equation}
\Delta^{\rm H}=-k r^4+r^2-2\widetilde{m}r+a^2 .
\end{equation}
Equation~(\ref{rdot_Weyl}) determines the radial evolution of photon trajectories in the Weyl BH spacetime.
{The photon dynamics considered in this work is governed by the full set of geodesic equations, which form a higher-dimensional Hamiltonian system. Although the radial equation is often presented separately for convenience, it represents a reduced description of the underlying phase space rather than a genuinely one-dimensional system. In particular, the motion involves variables such as $(r,\dot{r},\phi)$ (and, in general, additional angular degrees of freedom), ensuring that the system satisfies the dimensional requirements for complex dynamical behavior. Consequently, the Poincaré--Bendixson theorem is not applicable in this context.
In general, photon trajectories correspond to an open scattering problem. However, the present analysis focuses on regions of phase space near the photon sphere and quasi-bound orbits, where trajectories remain effectively confined over finite evolution times and exhibit strong sensitivity to initial conditions. In this regime, the use of phase-space diagnostics such as Poincaré sections and finite-time instability indicators provides meaningful insight into the dynamical structure of the system.
A fully rigorous treatment of unbounded motion would require the framework of chaotic scattering theory. The present study, however, is primarily concerned with characterizing the onset of irregular behavior and the transition in phase-space structure induced by rotation and conformal gravity corrections, rather than providing a complete scattering-theoretic description.}
%%%%%__________________________________SECTION_____________________________________________%%%%
\section{Null circular orbits and Lyapunov stability in Weyl gravity}\label{S3}
%%%%%______________________________________________________________________________________%%%%
Null circular orbits correspond to photon trajectories confined to a constant radius \(r=r_c\). These orbits satisfy the conditions
\begin{equation}
\dot r=0, \qquad \frac{d}{dr}(\dot r^2)=0.
\end{equation}
Defining the radial potential
\begin{equation}
R(r)=(E(r^2+a^2)-aL)^2-\Delta^{\rm H}(L-aE)^2 ,
\end{equation}
the circular orbit conditions reduce to
\begin{equation}
R(r_c)=0, \qquad R'(r_c)=0.
\end{equation}
Solving these equations, the critical impact parameter \(b=L/E\) is obtained as
\begin{equation}
b_\pm(r_c)=\frac{a\left(2\widetilde{m}r_c+k r_c^4\right)\pm r_c\sqrt{\Delta_c^{\rm H}\left(2\widetilde{m}r_c+k r_c^4\right)}}{2\widetilde{m}r_c+k r_c^4-r_c^2},
\end{equation}
where \(\Delta_c^{\rm H}=\Delta^{\rm H}(r_c)\) and the signs \(\pm\) correspond to retrograde and prograde photon orbits.
The radius of the photon sphere is determined from
\begin{equation}
r_c^2-3\widetilde{m}r_c \pm 2a\sqrt{\widetilde{m}r_c} + k r_c^4=0 .
\label{photon_Weyl}
\end{equation}
The effective potential for null geodesics is defined as
\begin{equation}
V_{\rm eff}(r)=-\frac{1}{r^4}\left[(E(r^2+a^2)-aL)^2-\Delta^{\rm H}(L-aE)^2\right].
\end{equation}
The instability of null circular orbits is characterized by the Lyapunov exponent, which measures the exponential divergence of nearby photon trajectories. Expanding the radial potential around \(r=r_c\) up to second order gives
\begin{equation}
R(r)\simeq \frac12 R''(r_c)(r-r_c)^2 .
\end{equation}
Substituting this expansion into the definition of the Lyapunov exponent yields
\begin{equation}
\lambda_{\rm null}=
\sqrt{\frac{R''(r_c)}{2\,\dot t_c^{\,2}\,r_c^{4}}}.
\end{equation}
Using the explicit form of \(\dot t_c\), the Lyapunov exponent for the Weyl BH can be written as
\begin{equation}
\lambda_{\rm null}=
\sqrt{
\frac{2\Delta_c^{\rm H}-r_c^2(\Delta_c^{\rm H})''}
{2r_c^2(r_c^2+a^2-a b_c)^2}
},
\label{lambda_Weyl}
\end{equation}
where
\begin{equation}
(\Delta_c^{\rm H})''=\left.\frac{d^2\Delta^{\rm H}}{dr^2}\right|_{r=r_c}.
\end{equation}
The positive Lyapunov exponent indicates that the null circular orbit is dynamically unstable. Larger values of \(\lambda_{\rm null}\) correspond to stronger sensitivity to perturbations in the initial conditions and enhanced chaotic behavior of photon trajectories.

%%%%%__________________________________SECTION_____________________________________________%%%%
\section{Poincaré Section}\label{S4}
%%%%%______________________________________________________________________________________%%%%
Poincaré sections, or Poincaré maps, take a reduced representation of the phase space dynamics, and clarify the study of non-linear dynamical systems and chaos. Closed, ordered curves result from regular or quasi-periodic orbits, and chaotic trajectories cause scattered and irregular distributions of points. Poincaré sections in BH spacetimes clarify how strong gravitational fields govern photons, illustrated in stable versus unstable and chaotic orbits. They are a mathematical-physical connector between phase space and how photons behave in strong gravity.

{
The equations of motion are solved numerically by rewriting them as a system of first-order differential equations for the radial coordinate, its conjugate velocity, and the azimuthal angle. The radial effective potential and its derivative are evaluated numerically, with the derivative computed using a second-order central finite-difference scheme with a small step size to ensure accuracy.
The system is integrated using an adaptive step-size Runge--Kutta method (RK45) as implemented in \texttt{scipy.integrate.solve\_ivp}. The relative and absolute tolerances are set to $10^{-7}$ and $10^{-9}$, respectively. The integration is performed over a sufficiently long time interval to capture the asymptotic behavior of trajectories, while the adaptive time-stepping automatically refines the step size in regions of rapid variation.
To ensure the reliability of the numerical results, we monitor the conservation of the effective Hamiltonian (equivalently, the radial constraint derived from the potential). The relative numerical error remains below $10^{-6}$ throughout the integration, confirming the stability and accuracy of the method.
Poincar\'e sections are constructed by recording phase-space points whenever the trajectory crosses the surface $\phi = \pi/2$ in the forward direction. The crossing points are determined using linear interpolation between successive integration steps, which improves the precision of the phase-space mapping.
To explore the global structure of the phase space, a grid of initial conditions is employed over a specified range of the radial coordinate and its velocity. 
}

\begin{figure*}
	\centerline{
		\includegraphics[width=170mm,height=120mm]{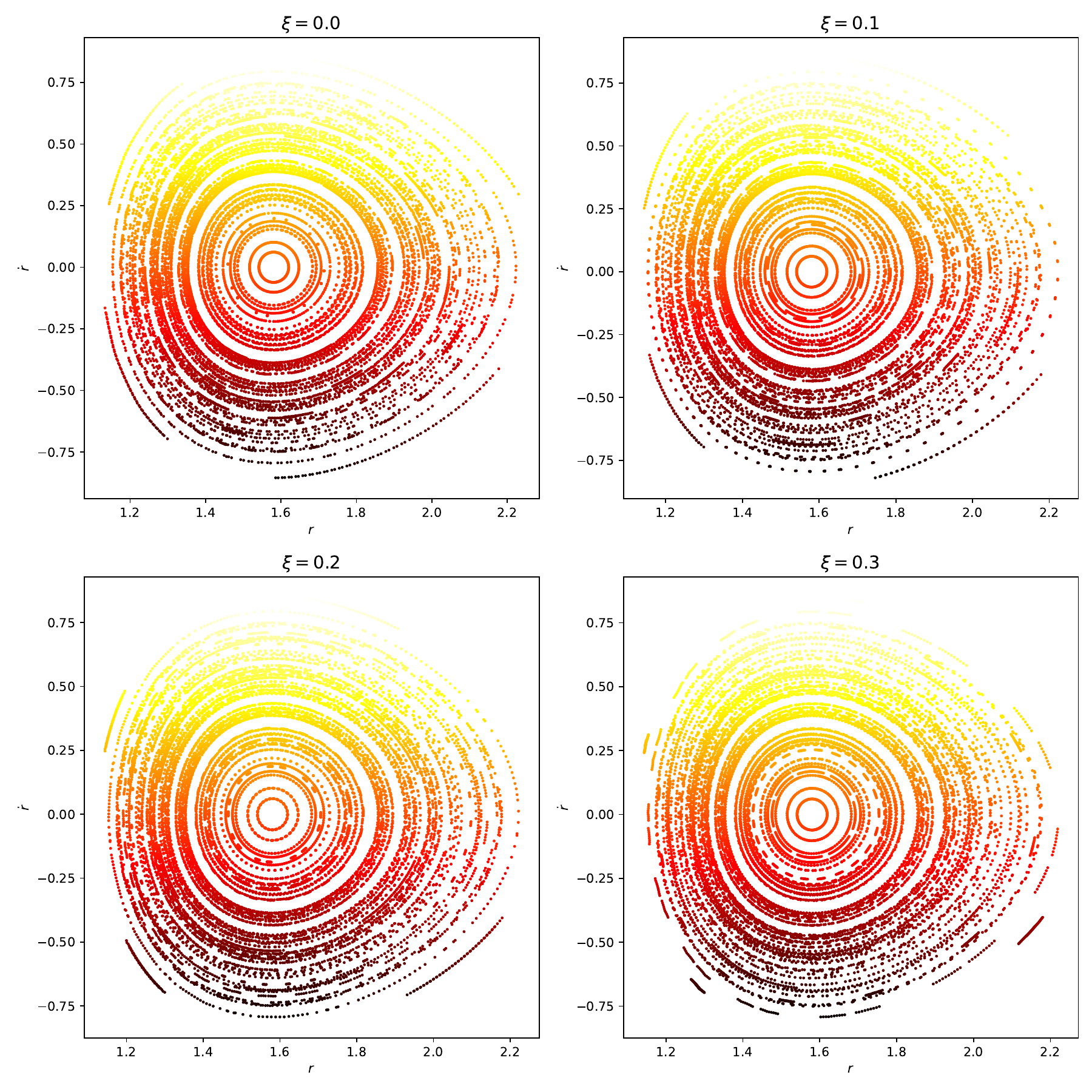}}
	\caption{Poincar\'e sections of null geodesic motion of Weyl rotating BHs at with $p_{\theta}>0$, for different values of the parameters $\xi$, obtained using various initial conditions. Throughout the plots, we fix $\kappa=0.01$, $\gamma=0.2$, $E=0.95$, and $L=3$.}
\label{fig1.1}
\end{figure*}

\begin{figure*}
	\centerline{
		\includegraphics[width=170mm,height=120mm]{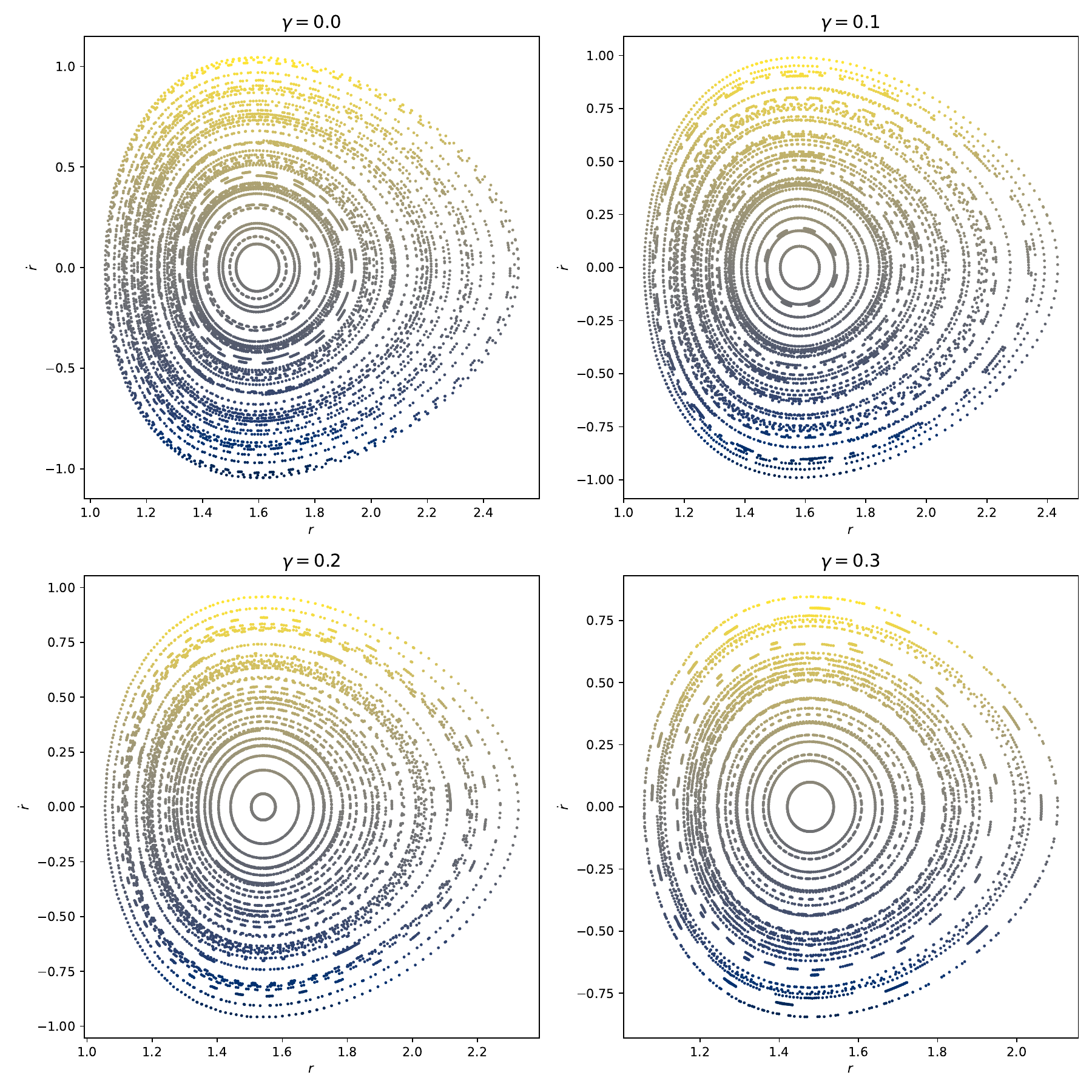}}
	\caption{Poincar\'e sections of null geodesic motion of Weyl rotating BHs at with $p_{\theta}>0$, for different values of the parameters $\gamma$, obtained using various initial conditions. Throughout the plots, we fix $\kappa=0.01$, $\xi=0.1$, $E=0.95$, and $L=3$.}
\label{fig2}
\end{figure*}

\begin{figure*}
	\centerline{
		\includegraphics[width=170mm,height=100mm]{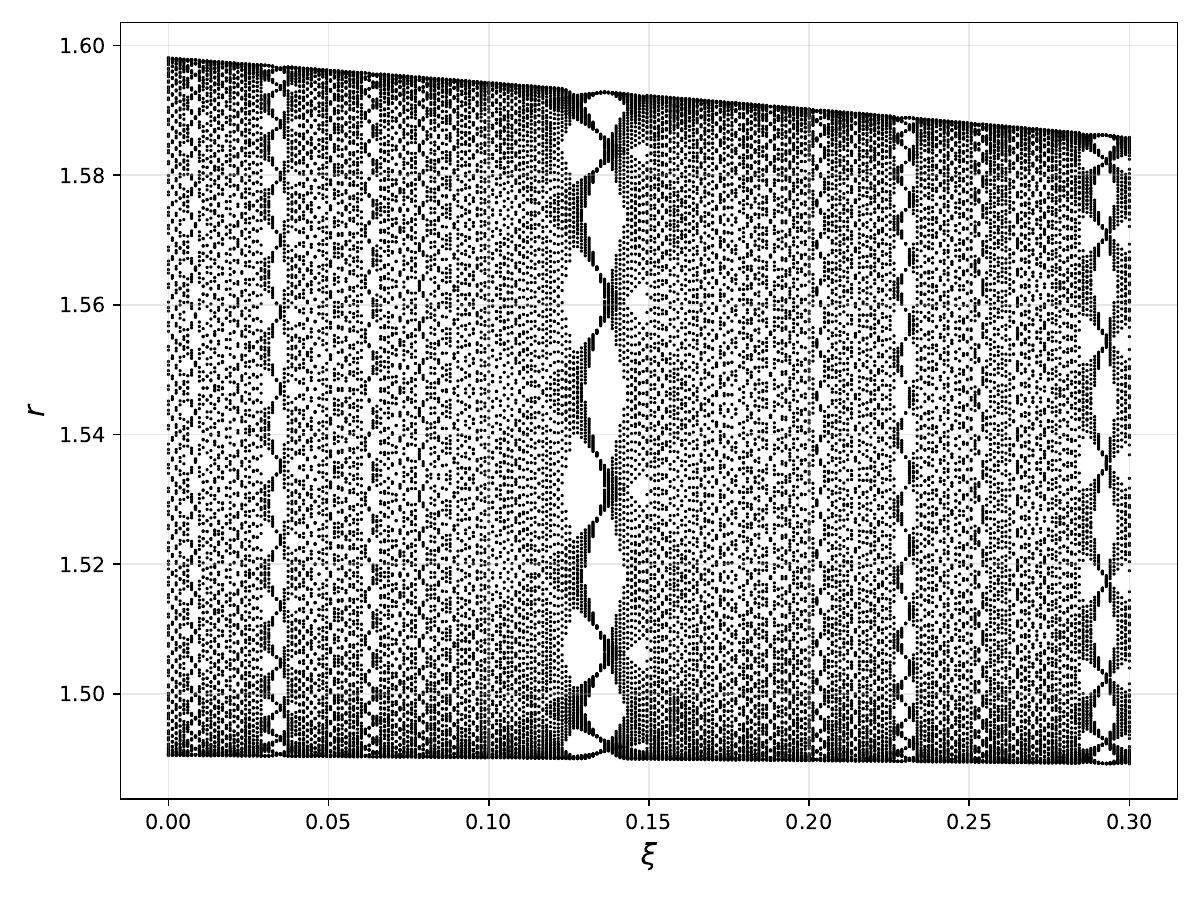}}
	\caption{{Bifurcation diagram of the Poincar\'e-section crossing radius $r$ versus the Weyl gravity parameter $\xi$. Here, we fix $\kappa=0.01$, $\gamma=0.2$, $E=0.95$, and $L=3$.}}
\label{figBifrcation}
\end{figure*}

{In Fig.~\ref{fig1.1}, Poincar\'e sections depict the phase-space structure of null geodesic motion for varying values of the parameter $\xi$, constructed over a representative set of initial conditions. For $\xi = 0$, the trajectories form well-defined, closed invariant curves, indicating predominantly regular and integrable motion. As $\xi$ increases, these invariant tori gradually deform, reflecting the growing influence of the modified gravitational background. At intermediate values ($\xi = 0.1, 0.2$), slight irregularities and dispersion of points begin to appear, suggesting the onset of weak chaos. For larger values ($\xi = 0.3$), the phase-space structure exhibits further distortion with partial breakdown of invariant curves, indicating enhanced nonlinear behavior.}
Fig.~\ref{fig2} presents the Poincar\'e sections of null geodesic motion for varying values of the parameter $\gamma$, with $\kappa = 0.01$, $\xi = 0.1$, $E = 0.95$, and $L = 3$ held fixed throughout. Across all four panels, the phase space is dominated by smooth, nested invariant curves, indicating that the null geodesic motion remains predominantly regular and integrable in this parameter regime. For $\gamma = 0.0$, the trajectories trace concentric, well-ordered closed loops spanning a broad radial range, with no visible signature of chaotic dispersion. As $\gamma$ increases to $0.1$ and $0.2$, the qualitative structure is preserved. The invariant curves remain closed and orderly, while the accessible radial extent of the section contracts slightly and the density of the nested rings increases marginally near the inner boundary. At $\gamma = 0.3$, the outer envelope of the phase-space structure contracts further still, compressing the invariant curves into a narrower radial band, though the underlying regularity of the orbits is not qualitatively disrupted. Overall, increasing $\gamma$ acts primarily to reshape and compress the effective radial domain of the null geodesic phase space rather than to induce chaotic behavior, in contrast to the destabilizing effect of increasing $\xi$ seen in Fig.~\ref{fig1.1}. {It should be noted that, due to computational limitations, the phase-space sampling is finite and the integration time is restricted; therefore, the fine-grained structure of chaotic regions may not be fully resolved. Nevertheless, the overall trend clearly demonstrates a transition from regular to increasingly complex dynamics with increasing $\xi$.}
\\
{To assure the nature of the dynamical behavior inferred from the Poincaré sections, we have plotted the bifurcation diagram shown in Fig.~\ref{figBifrcation} by recording the Poincaré-section crossing radius as the Weyl gravity parameter $\xi$ is varied. The diagram exhibits a bounded and highly organized distribution of intersection points, with smooth upper and lower envelopes throughout the investigated parameter range. No period-doubling cascade, branch proliferation, or irregular broadening of the occupied region is observed. Further, the crossing points remain confined within a well-defined band, indicating that the underlying trajectories preserve their ordered character. The fact that the crossing points remain inside a specific band shows that the underlying trajectories continue to be ordered. It is possible to interpret the small gaps at specific parameter values as resonance structures inside the regular phase space. The bifurcation diagram indicates that the orbital dynamics are mostly quasi-periodic overall. Rather than going to complete chaos, the motion evolves on invariant tori.}

%%%%%__________________________________SECTION_____________________________________________%%%%
\section{Lyapunov Stability Indicators}\label{S5}
Identification and quantification of instability of relativistic photon orbits near compact astrophysical objects is made possible by applying Lyapunov exponents to characterize non-linear dynamical systems and determine how divergent the trajectories of the system are from one another in the phase space \cite{pesin1977characteristic,benettin1980lyapunov}, {while a modified neural network method for computing the Lyapunov exponent spectrum in the nonlinear analysis of dynamical systems studied in \cite{yakovleva2025modified}.} In this context, a positive Lyapunov exponent is interpreted to reflect a chaotic motion, whereas a negative exponent is settled and a zero exponent is indicative of being marginally settled or quasi-periodic. In this paper, we use two well-established Lyapunov-based measures, the LLE and the FLI, to help identify and categorize these types of behavior. 
{In relativistic dynamical systems, the characterization of chaos through Lyapunov exponents requires careful consideration of coordinate invariance and covariant formulations, and it established that physically meaningful measures of chaos must be independent of the choice of coordinates, ensuring consistency in curved spacetime analyses~\cite{Motter2003,Xu2003,Cornish2003}. Recent developments further highlight the role of Lyapunov exponents as  powerful probes of gravitational dynamics and potential deviations from GR, particularly in strong field regimes~\cite{Deich2024} which provide a solid theoretical foundation for employing Lyapunov based indicators in the analysis of relativistic chaotic motion.}

%%%%%__________________________________SUBSECTION_____________________________________________%%%%
\subsection{Largest Lyapunov Exponent}
%%%%%______________________________________________________________________________________%%%%

The LLE characterizes how sensitive photon paths are to changes in their starting conditions and how quickly nearby paths in phase space diverge. When strong gravitational fields are present, photon paths are null geodesics and demonstrate chaotic behavior. Let $\tau$ represent affine parameter and $d(\tau)$ represent the distance between two nearby photon paths. The LLE is defined as~\cite{Wu:2006rx,Han:2008zzf,Chen:2016tmr},
\begin{equation}
\lambda = \lim_{\tau \to \infty} \frac{1}{\tau} \ln\!\left(\frac{d(\tau)}{d(0)}\right),
\end{equation}
where $d(0)$ represents the initial separation. This covariant formulation guarantees that $\lambda$ is independent of the choice of coordinates and therefore provides a physically meaningful measure of orbital instability in curved spacetime. {As we are dealing with null geodesics, the notion of a physical time parameter is not directly applicable, unlike the case of timelike geodesics. The proper time vanishes identically along null trajectories and therefore cannot serve as a valid evolution parameter for photon motion. In this work, the Lyapunov exponent is formulated using an affine parameter along the null geodesics, ensuring a geometrically consistent description of the dynamics. Consequently, the Lyapunov exponent defined in this framework is invariant under coordinate transformations and does not depend on the choice of spacetime coordinates.}

We focus on the stability analysis of the null circular geodesics and expand the radial equation of motion about a reference circular orbit at $r = r_0$, $r = r_0 + \delta r$ with a small perturbation. The perturbation $\delta r$ is a variable that captures the response of the radial null geodesics and is governed by a linearized equation that encapsulates the behavior of radial null geodesics. When there is affine parameter $\delta r$ and it increases exponentially, the orbit is deemed to be unstable and is said to have extreme exponential divergence. In such an event, the sensitivity to the initial conditions increases, thus showing chaotic motion. On the other hand, If, additionally, $\delta r$ is exponentially bounded or $\delta r$ diminishes, it then implies that there is regular motion in the circular orbit which is then said to be stable and the irregular unstable initial conditions of perturbations are also said to be then dynamically stable. Physically, the circular geodesics of a sphere with a BH has extremely destabilized geometry. Statistically, the layout of the BH is an extreme collapse of the geometry of the standard elliptic orbit, which has a catastrophic ending, and with extremely rapid exponential behaviour of the confined chaotic motion of the orbits. This illustrates that the Lyapunov exponent is a strong diagnostic that describes the regular to chaotic transition in the motion of photons around rotating BHs.
{To ensure that the perturbative analysis remains well-defined and free from numerical or coordinate singularities, we impose a strict bound on the radial perturbation parameter $\delta r$. Since the motion must remain outside the event horizon, determined by $\Delta^{\mathrm{H}}(r)=0$, we require
\begin{equation}
r_0 + \delta r > r_h,
\end{equation}
where $r_h$ is the largest real root of $\Delta^{\mathrm{H}}(r)$. This leads to the constraint
\begin{equation}
|\delta r| < r_0 - r_h.
\end{equation}
Furthermore, to preserve the validity of the linearized expansion around the circular null geodesic, the perturbation must remain sufficiently small. Combining these requirements, we adopt the bound
\begin{equation}
|\delta r| \leq \delta_{\max} = \min\left( r_0 - r_h,\; \left|\frac{V_{\text{eff}}'(r_0)}{V_{\text{eff}}''(r_0)}\right| \right),
\end{equation}
which ensures that the trajectory remains within the regular domain of the spacetime and that the Lyapunov analysis remains valid.}

\begin{figure*}
	\centerline{
		\includegraphics[width=170mm,height=100mm]{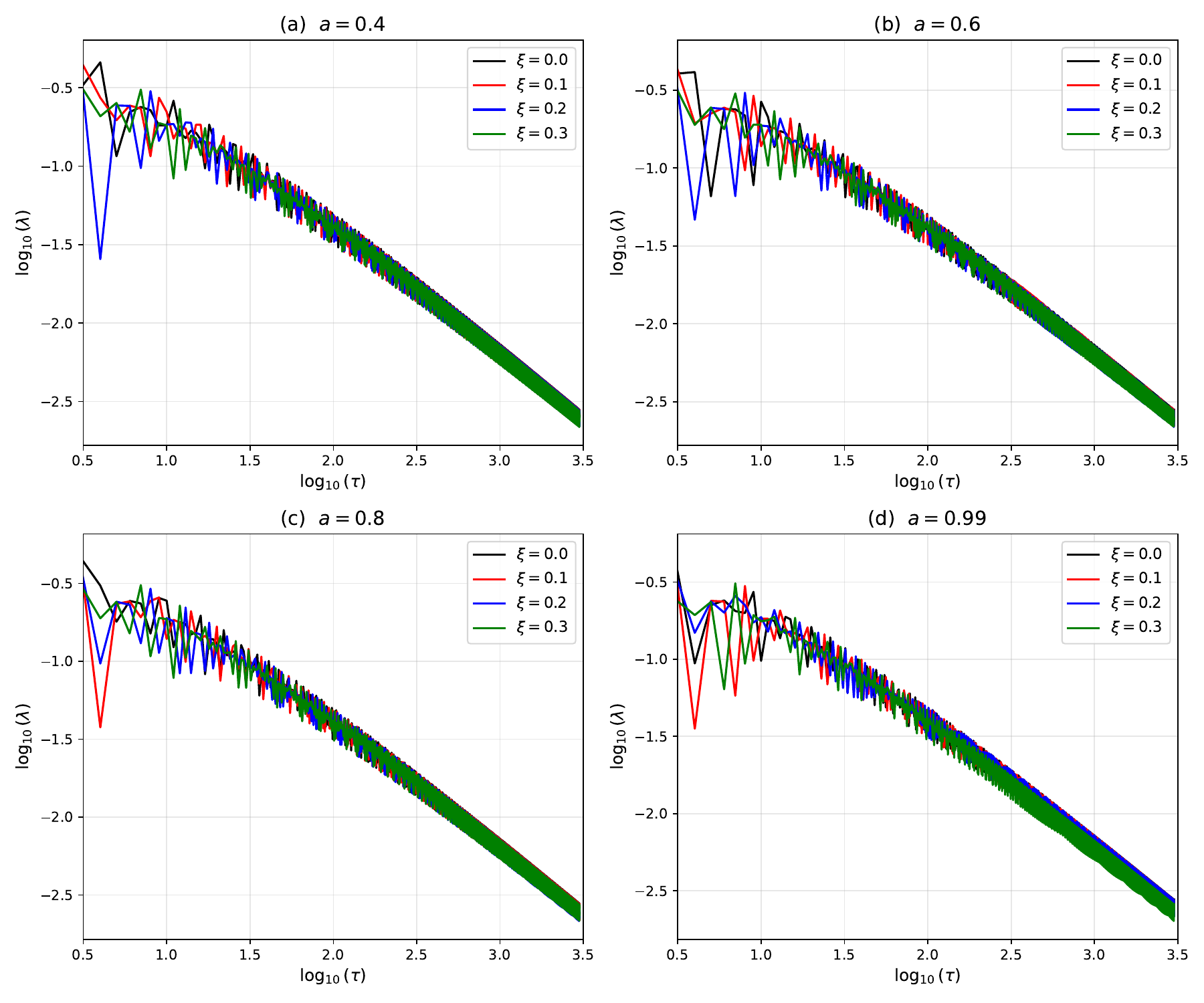}}
	\caption{Logarithmic Lyapunov exponent $\log_{10}\lambda$ versus $\log_{10}\tau$ for rotating BHs in WCG, shown for different values of the parameter $\xi$. Panels (a)-(d) correspond to increasing values of the rotation parameter $a$. The remaining parameters are fixed at $\kappa = 0.01$, $\gamma = 0.2$, $L = 3$, and $r_0 = 1.2$.}
\label{fig4}
\end{figure*}

\begin{figure*}
	\centerline{
		\includegraphics[width=170mm,height=100mm]{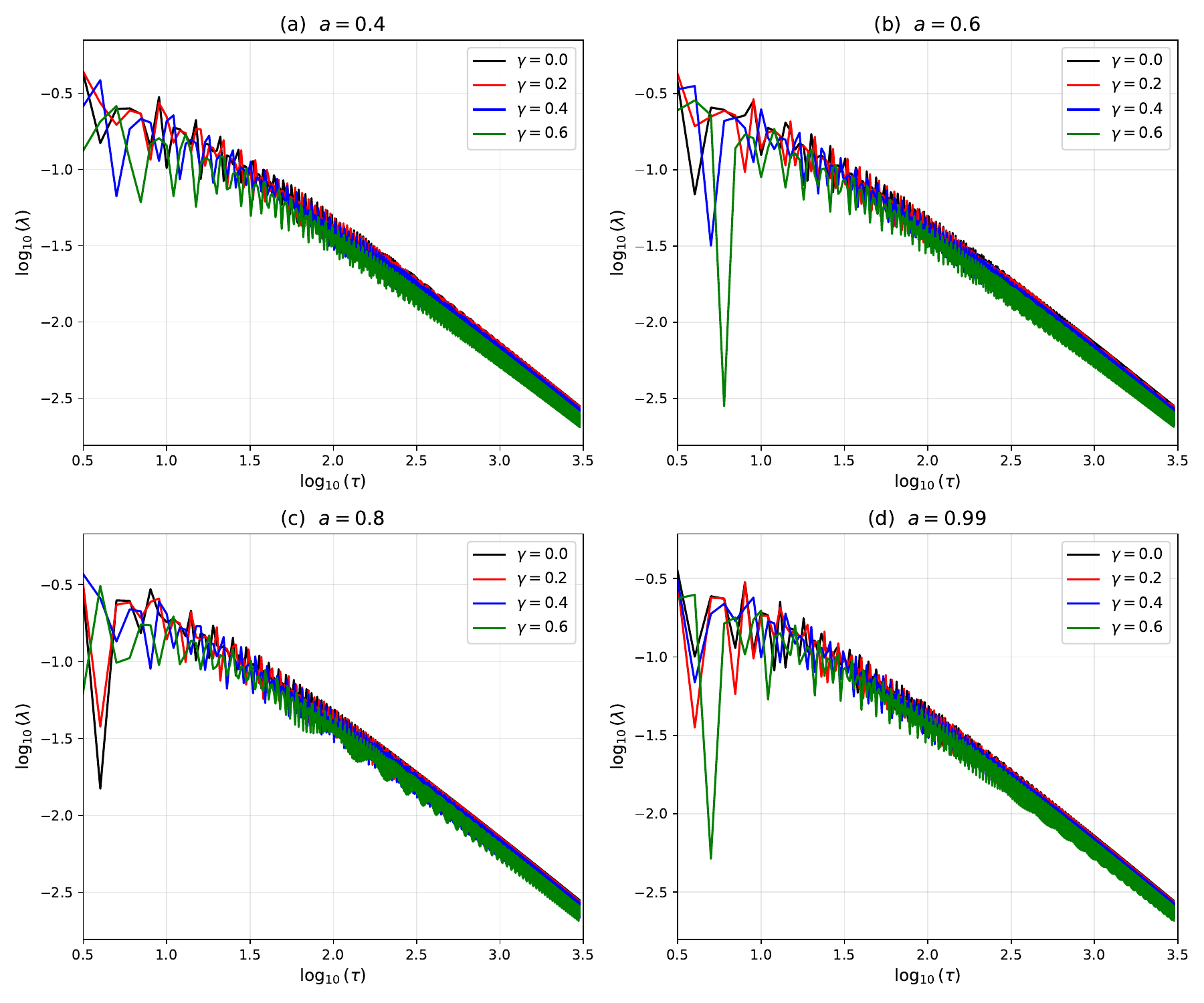}}
	\caption{Logarithmic Lyapunov exponent $\log_{10}\lambda$ versus $\log_{10}\tau$ for rotating BHs in WCG, shown for different values of the parameter $\gamma$. Panels (a)-(d) correspond to increasing values of the rotation parameter $a$. The remaining parameters are fixed at $\kappa = 0.01$, $\xi = 0.1$, $L = 3$, and $r_0 = 1.2$.}
\label{fig5}
\end{figure*}

Figs.~\ref{fig4} and~\ref{fig5} present the evolution of the logarithmic Lyapunov exponent, $\log_{10}\lambda$, as a function of the logarithmic affine parameter, $\log_{10}\tau$, for photon trajectories around rotating BHs in WCG. These figures provide a quantitative measure of orbital sensitivity and stability and allow us to characterize the transient dynamical response of the photon trajectories. Fig.~\ref{fig4} illustrates the dependence of the Lyapunov exponent on the dimensionless mass-dependent parameter $\xi$ and the rotation parameter $a$. The curves exhibit pronounced transient oscillations at small values of $\tau$, reflecting short-time sensitivity to initial conditions and the influence of the local phase-space structure. Further, as $\tau$ increases, the Lyapunov exponent gradually approaches a monotonic decay, indicating convergence toward the asymptotic dynamical regime. The increasing separation between the curves corresponding to different values of $\xi$ with larger $a$ demonstrates that mass-dependent conformal corrections modify the transient stability properties of photon orbits. Physically, this behavior suggests that the deformation of the effective photon potential induced by $\xi$ can enhance the short-time sensitivity of photon trajectories in the vicinity of the photon sphere.

In Fig.~\ref{fig5}, the parameter $\gamma$, which arises from the fourth-order nature of WCG, is varied across the panels, while the rotation parameter $a$ increases from panel (a) to panel (d) and all remaining parameters are held fixed. The magnitude of the logarithmic Lyapunov exponent and the amplitude of its transient variations increase systematically with increasing $\gamma$, indicating enhanced short-time divergence and orbital sensitivity. Notably, for higher rotation cases, a sudden decrease in the Lyapunov exponent is observed at early times for specific values of $\gamma$. The temporary suppression of this divergence may arise from a balance between frame-dragging effects and conformal-curvature corrections, leading to transiently more stable configurations of photon trajectories. However, these transient features do not imply a sustained positive Lyapunov exponent or fully developed chaotic dynamics. At late times, the Lyapunov exponent decreases toward zero, consistent with the regular character of the reduced equatorial dynamics.

Therefore, the effects of $\xi$ and $\gamma$ primarily modify the transient instability and sensitivity of null geodesics rather than establishing a definitive transition to chaos. The observed behavior provides a quantitative measure of how the higher-order curvature corrections and rotation influence the stability of photon trajectories in rotating WCG black holes.

%__________________________________SUBSECTION_____________________________________________%%%%
\subsection{Fast Lyapunov Indicator}
%%%%%______________________________________________________________________________________%%%%

FLI provides a robust additional diagnostic for the Poincaré section by allowing an orbital stability characterization in phase space. {In contrast to classical Lyapunov exponents, which require long integration times to converge, the FLI is a finite-time indicator that distinguishes regular from chaotic orbits over relatively short evolution intervals.}

Let $d(\tau)$ denote the norm of the deviation vector between two initially nearby geodesics at affine parameter $\tau$, with $d(0)$ representing their initial separation. The FLI is defined as~\cite{Wu:2006rx,Han:2008zzf,Chen:2016tmr},
\begin{equation}
\mathrm{FLI}(\tau) = \log_{10}\!\left(\frac{d(\tau)}{d(0)}\right).
\end{equation}
{The evolution of the deviation vector is governed by the variational equations derived from the geodesic equations, ensuring that the FLI captures the local dynamical response of photon trajectories to spacetime curvature.}
For regular or quasiperiodic motion, the time-along $\mathrm{FLI}(\tau)$ almost linearly increases and there is a slow increase when the affine parameter is used as a reference. {In contrast, trajectories exhibiting strong sensitivity to initial conditions show a much faster growth of the FLI, indicating enhanced local instability.} The difference in behaviour is large enough that the time evolution of the FLI provides a {useful} method of {characterizing} the dynamical behavior.
{It is important to note that the FLI is a finite-time diagnostic and does not correspond to the asymptotic Lyapunov exponent. Rather, it provides a qualitative measure of transient dynamical instability, which is particularly useful in systems where long-time bounded motion is not guaranteed.}

\begin{figure*}
	\centerline{
		\includegraphics[width=170mm,height=100mm]{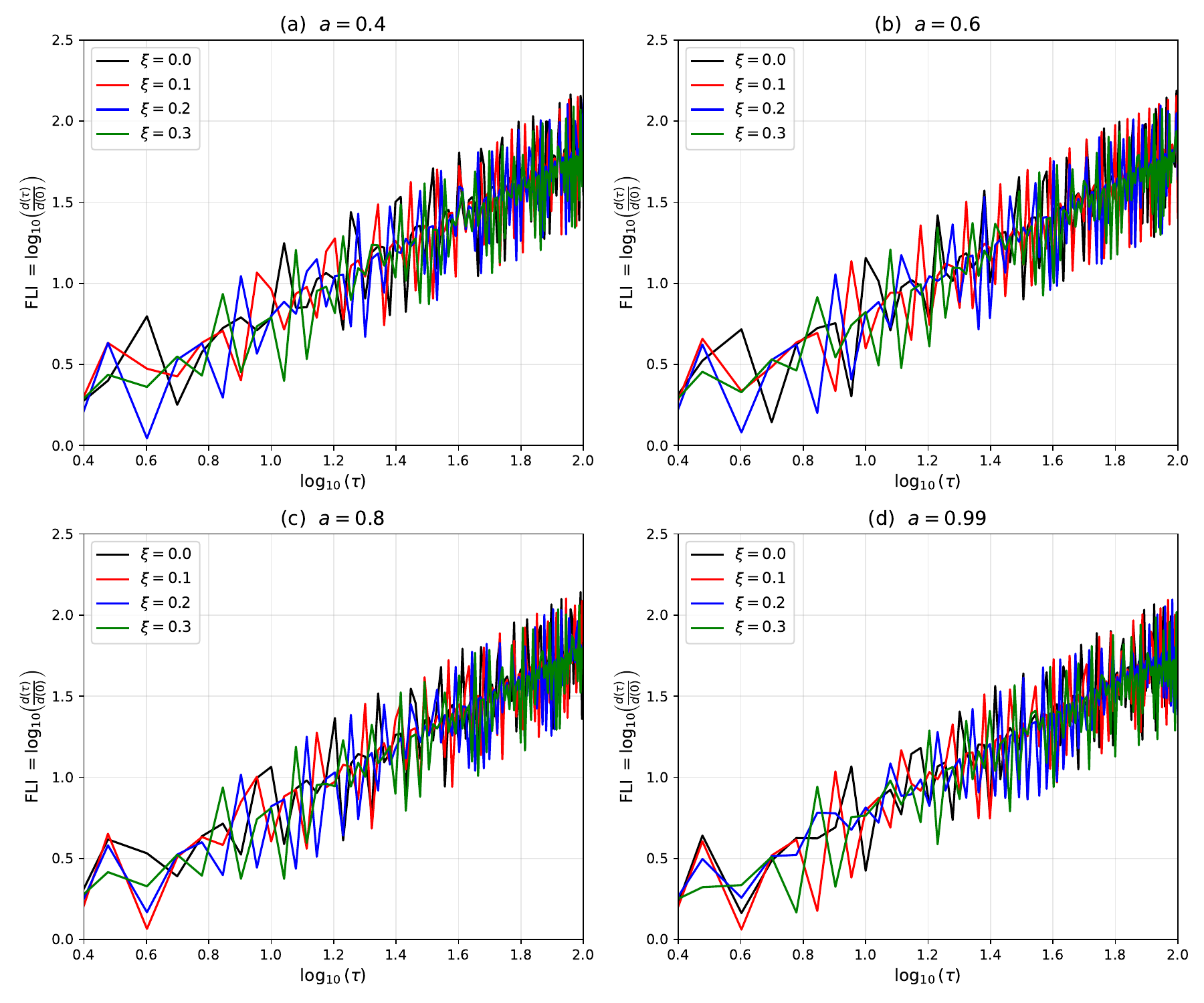}}
	\caption{The FLI versus affine parameter for rotating BHs in WCG for various values of the parameter $\xi$. From panel (a) to panel (d), the parameter sets $a$ are chosen to be increasing. The remaining parameters are fixed at $\kappa = 0.01$, $\gamma = 0.2$, $L = 3$, and $r_0 = 1.2$. }
\label{fig6}
\end{figure*}

\begin{figure*}
	\centerline{
		\includegraphics[width=170mm,height=100mm]{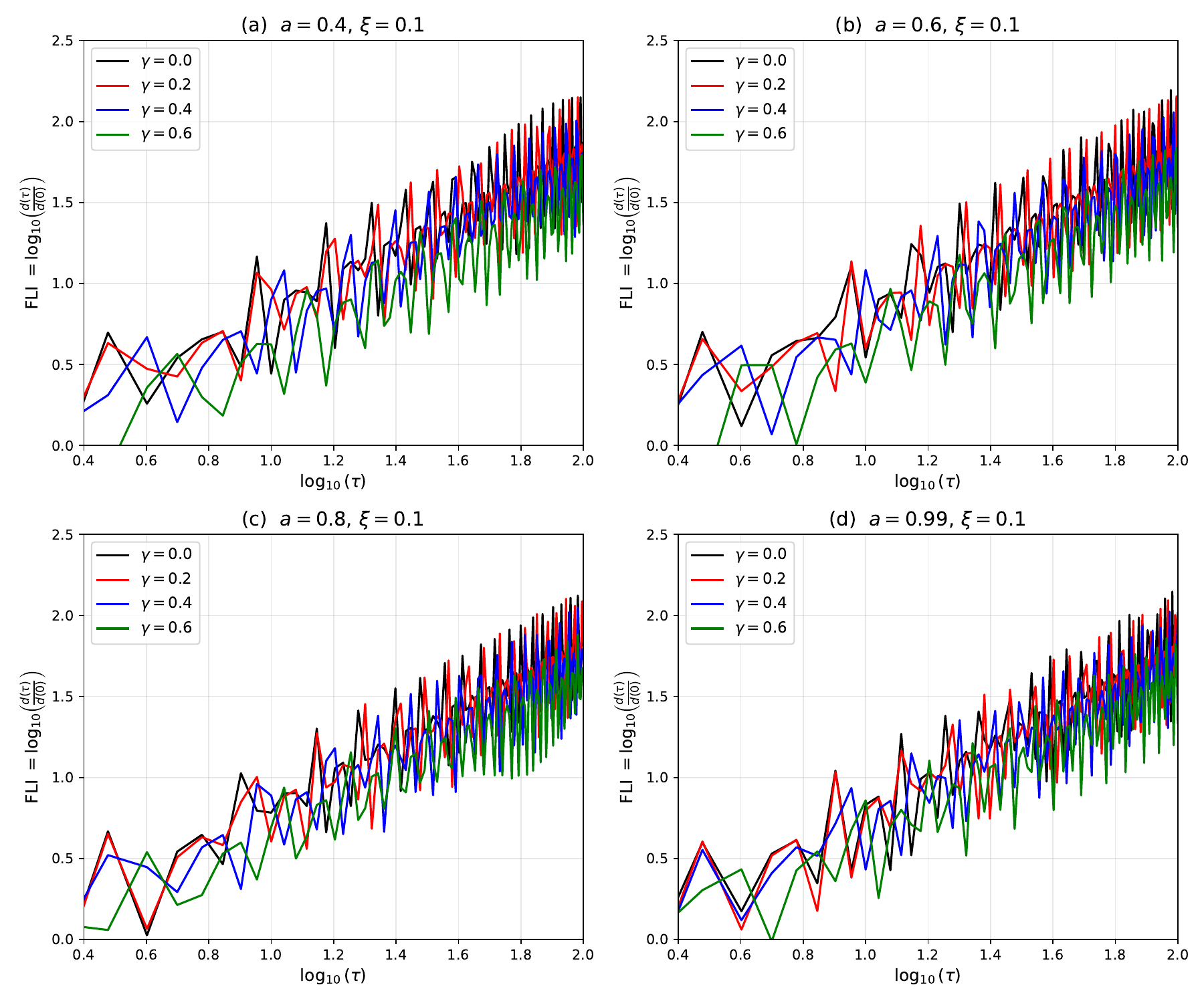}}
	\caption{The FLI versus affine parameter for rotating BHs in WCG for various values of the parameter $\gamma$. From panel (a) to panel (d), the parameter sets $a$ are chosen to be increasing. The remaining parameters are fixed at $\kappa = 0.01$, $\xi = 0.1$, $L = 3$, and $r_0 = 1.2$. }
\label{fig7}
\end{figure*}

In Figs. \ref{fig6} and \ref{fig7}, we analyze the temporal evolution of the FLI for rotating WCG black holes for various values of the dimensionless mass-dependent parameter $\xi$ and the conformal-gravity parameter $\gamma$. The FLI is sensitive to the divergence of initially neighboring trajectories and therefore provides a useful diagnostic for characterizing the dynamical behavior of photon orbits in the strong-gravity regime. In all cases, the FLI exhibits an overall increasing tendency accompanied by pronounced oscillations. However, this increase should not be interpreted by itself as evidence of fully developed chaotic dynamics.

As $\xi$ increases, the amplitude and growth rate of the FLI change systematically, indicating a stronger sensitivity of the photon trajectories to the underlying gravitational parameters. The oscillatory behavior reflects variations in the local instability of the photon trajectories rather than providing, by itself, a definitive signature of chaos. Figure \ref{fig7} further illustrates the dependence of the FLI on $\xi$ and $\gamma$ for the rotating WCG geometry. As $\gamma$ increases, the FLI curves exhibit a more pronounced growth, indicating enhanced orbital sensitivity in the corresponding parameter regime.

The quantitative analysis of the FLI curves, including the fitted growth trends, shows that the increase remains non-exponential and is consistent with regular or weakly unstable dynamics rather than a clear signature of sustained chaos. Accordingly, large FLI values or increasing FLI curves alone are not taken as sufficient evidence for chaotic photon motion. The parameters $\xi$ and $\gamma$ nevertheless play an important role in modifying the stability and predictability of photon trajectories around rotating WCG black holes. Thus, the FLI analysis provides a quantitative characterization of orbital sensitivity and its dependence on the conformal-gravity parameters, without implying a transition to fully chaotic dynamics based solely on the observed FLI growth.

%%%%%__________________________________SUBSECTION_____________________________________________%%%%
\section{Kolmogorov--Sinai entropy}\label{S6}
%%%%%______________________________________________________________________________________%%%%
The Kolmogorov--Sinai (KS) entropy provides a rigorous quantitative measure of chaoticity in a dynamical system, capturing the rate at which information about a trajectory is produced or lost over time in phase space. In Hamiltonian systems, the KS entropy is intimately connected to the Lyapunov spectrum via Pesin's theorem, which states that it equals the sum of all positive Lyapunov exponents~\cite{pesin1977characteristic},
\begin{equation}
h_{\mathrm{KS}} = \sum_{\lambda_i > 0} \lambda_i.
\end{equation}

{In the present work, we do not compute the asymptotic KS entropy in the strict mathematical sense. Instead, we analyze finite-time instability indicators derived from the divergence of nearby null geodesics~\cite{Pradhan:2015aaa,Mondal:2021exj}. These quantities provide a qualitative measure of dynamical instability over finite evolution times, particularly in regimes where long-time bounded motion is not guaranteed.}
{Accordingly, the quantity plotted and discussed in this section should be interpreted as a finite-time analogue of entropy-like behavior, rather than a direct evaluation of the KS entropy or a strict application of Pesin's identity.}

Physically, {these finite-time indicators} quantify the rate at which initially neighboring photon trajectories separate in the curved spacetime of a rotating BH in WCG. As such, {they provide a qualitative measure of the unpredictability} inherent in photon dynamics under strong gravitational fields. {Larger values correspond to enhanced local instability, while smaller values indicate more regular motion.} By computing these quantities across a range of initial conditions and BH parameters, one can map out regions of regularity and irregular behavior, thereby gaining insight into the structure and stability of photon orbits in this conformal gravity background.

\begin{figure*}
	\centerline{
		\includegraphics[width=170mm,height=220mm]{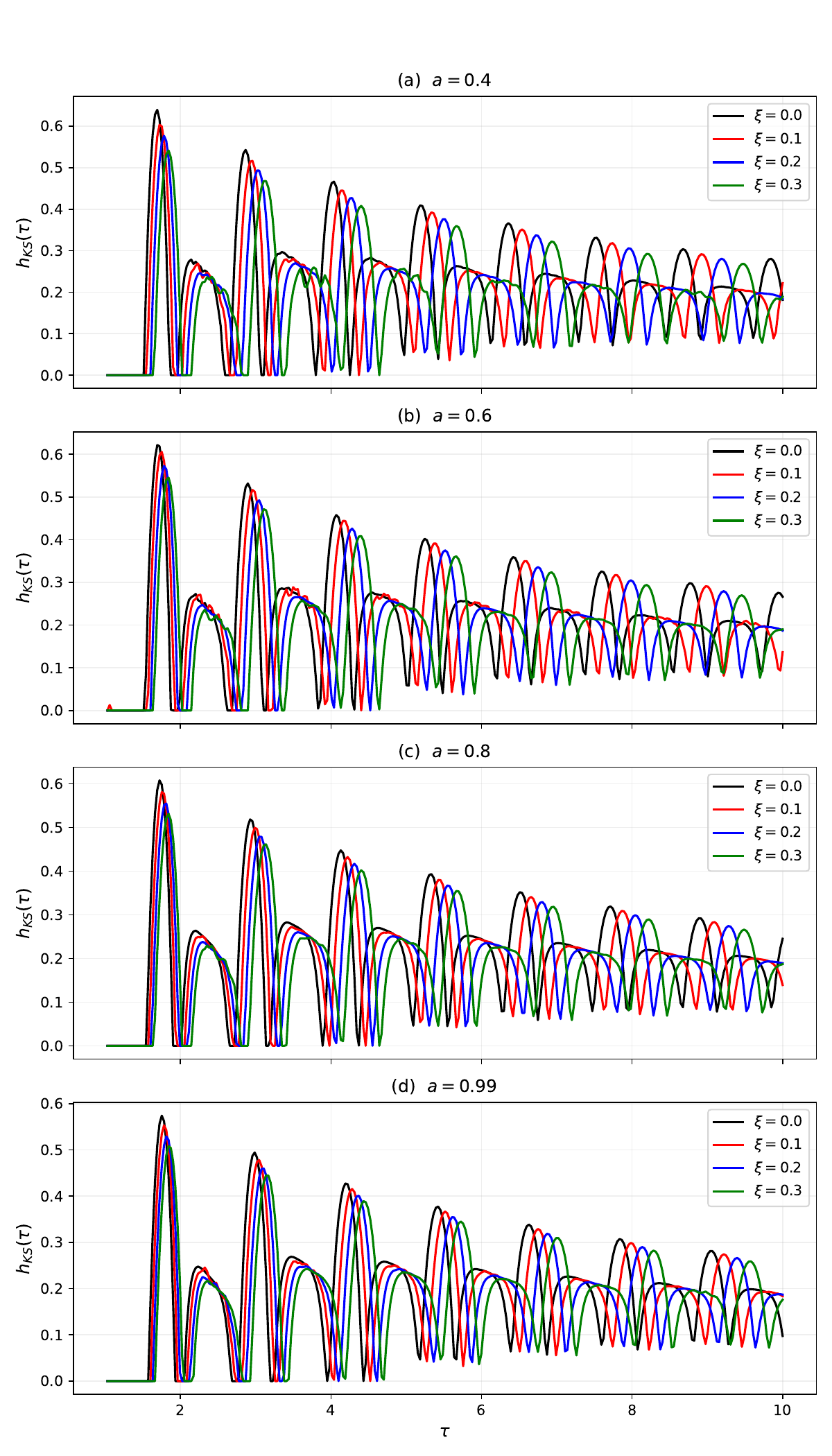}}
	\caption{{Time evolution of a finite-time instability measure (entropy-like quantity)} for null geodesics of rotating BHs in WCG for different values of the charge parameter $q$. The panels from top to bottom show increasing values of the rotation parameter $a$. The parameters are fixed at $\kappa = 0.01$, $\gamma = 0.2$ and $L = 3$.}
\label{fig8}
\end{figure*}

\begin{figure*}
	\centerline{
		\includegraphics[width=170mm,height=220mm]{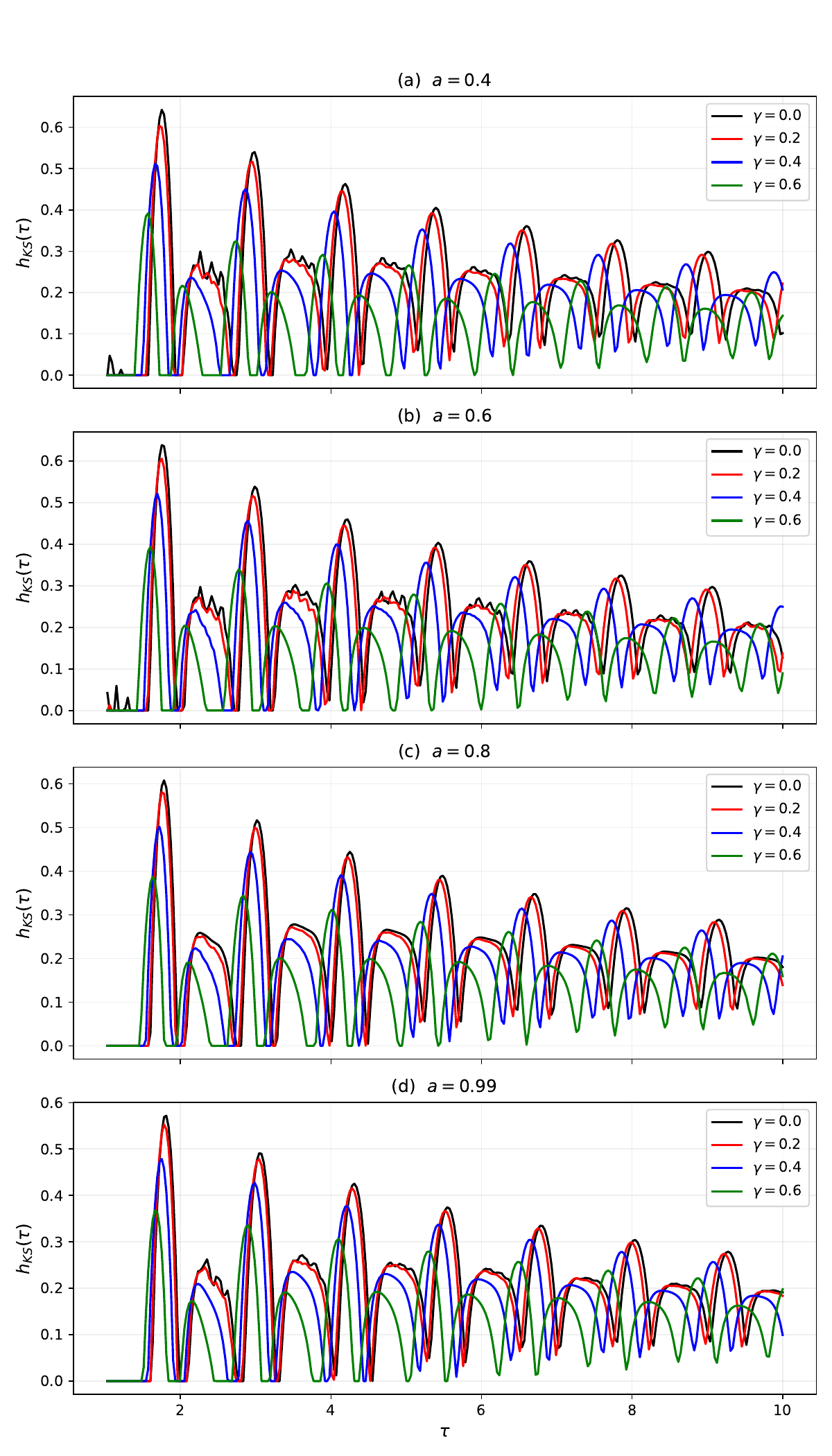}}
	\caption{{Time evolution of a finite-time instability measure (entropy-like quantity)} for null geodesics of rotating BHs in WCG for different values of the parameter $\gamma$. The panels from top to bottom show increasing values of the rotation parameter $a$. The parameters are fixed at $\kappa = 0.01$, $\xi = 0.1$ and $L = 3$.}
\label{fig9}
\end{figure*}

Fig.~\ref{fig8} shows the time evolution of {the finite-time instability measure} for null geodesics in rotating BHs in WCG for different values of the dimensionless mass-dependent parameter $\xi$ and increasing rotation parameter $a$ from panels (a) to (d). The remaining parameters are held fixed. The {measure} exhibits oscillatory behavior at early times, reflecting transient dynamical responses associated with strong-field photon motion and sensitivity to initial conditions. Further evolution in affine parameter reveals a gradual damping of oscillations and convergence toward a quasi-saturated regime, indicating the emergence of transient dynamical behavior. Moreover, increasing $\xi$ systematically modifies both the amplitude and frequency of oscillations, demonstrating that conformal mass-dependent corrections influence the instability of photon trajectories. Furthermore, the effect becomes more pronounced with increasing rotation parameter $a$. In particular, near extremal rotation $a=0.99$ leads to enhanced fluctuations and larger values of the measure over extended time intervals, indicating stronger local instability. The combined influence of rotation and the conformal parameter $\xi$ enhances nonlinear coupling in the geodesic equations and promotes the growth of irregular behavior associated with unstable photon orbits near the photon sphere.

The same analysis for different values of the parameter $\gamma$ is presented in Fig.~\ref{fig9}. The {finite-time instability measure} again displays oscillatory behavior at early times, indicating transient dynamical responses associated with strong-field photon motion and sensitivity to initial conditions. Further progression in affine parameter shows a gradual damping of oscillations and convergence toward a quasi-saturated regime. Moreover, variations in $\gamma$ systematically alter the amplitude and frequency of the oscillations, revealing the influence of higher-order curvature corrections on the instability of photon trajectories. Furthermore, the separation between the curves corresponding to different values of $\gamma$ is slightly larger than that observed for the parameter $\xi$, demonstrating a stronger sensitivity to conformal gravity effects encoded in $\gamma$. In particular, near extremal rotation $a=0.99$ produces enhanced fluctuations and higher values of the measure, indicating increased dynamical instability. The combined influence of BH rotation and the parameter $\gamma$ strengthens nonlinear coupling in the geodesic dynamics and enhances irregular behavior associated with unstable photon orbits.

{These results highlight the role of the parameters $\xi$ and $\gamma$, together with BH rotation, in influencing the degree of dynamical instability and loss of predictability in photon motion within the strong-gravity regime.}
{Strict equality between KS entropy and the sum of positive Lyapunov exponents holds under assumptions such as ergodicity (Pesin’s theorem), which are not explicitly verified here. Thus, the reported values should be interpreted as indicators of dynamical instability rather than exact entropy measures.}

%%%%%__________________________________SECTION_____________________________________________%%%%
\section{Quantitative Detection of Chaos Using Weighted Birkhoff Averages}\label{S7}
%%%%%______________________________________________________________________________________%%%%
Poincaré sections and Lyapunov exponents provide primarily qualitative diagnostics for the identification of chaotic dynamics. Here, we extend our analysis by employing the WBA as a robust numerical technique to discriminate between regular and chaotic photon trajectories in the spacetime of rotating BHs within WCG. This method exploits the convergence properties of time-averaged observables along individual orbits, thereby enabling an efficient classification of quasi-periodic versus chaotic motion~\cite{eckmann1985ergodic,cornfeld2012ergodic,das2016measuring,meiss2021birkhoff,sander2020birkhoff,duignan2023distinguishing}. 

For equatorial null geodesics ($\theta=\pi/2$) with $p_\theta>0$, the reduced phase space is characterized by the canonical coordinates $(r,p_r)$. Each photon orbit intersects a prescribed Poincaré surface of section multiple times, generating a discrete sequence of points that defines an associated iterated map
\begin{equation}
\mathbf{v}_{n+1} = \mathcal{M}(\mathbf{v}_n),
\end{equation}
where $\mathcal{M}$ is the Poincaré return map associated with the continuous geodesic flow.
Along this discrete orbit, we define a smooth observable
\begin{equation}
f(r,p_r) = \sin\!\big[2\pi(r+p_r)\big],
\end{equation}
which has been shown to effectively discriminate between different motion types in Hamiltonian systems. The weighted Birkhoff average of $f$ over $N$ successive iterations is given by~\cite{cornfeld2012ergodic,das2016measuring}
\begin{equation}
WB_N(f)(\mathbf{v}_0) = \sum_{n=0}^{N-1} w_{n,N} \, f\!\left(\mathcal{M}^n(\mathbf{v}_0)\right),
\end{equation}
with normalized weights
\begin{equation}
w_{n,N} = \frac{g(n/N)}{\sum_{k=0}^{N-1} g(k/N)}.
\end{equation}
The weighting function is constructed using a smooth bump profile~\cite{eckmann1985ergodic}:
\begin{equation}
g(z)=
\begin{cases}
\exp\!\left[-\dfrac{1}{z(1-z)}\right], & 0<z<1, \\
0, & \text{otherwise},
\end{cases}
\end{equation}
which mitigates boundary effects and accelerates convergence for regular trajectories, while chaotic orbits exhibit slower, irregular convergence. To quantify the convergence behavior, we compute two WBAs from successive segments of the same trajectory. Their difference defines a DIC~\cite{eckmann1985ergodic,cornfeld2012ergodic},
\begin{equation}
\mathrm{DIG} = -\log_{10}\!\left| WB_N^{(1)} - WB_N^{(2)} \right|.
\end{equation}
Large DIG values indicate rapid convergence and regular (quasi-periodic) trajectories, whereas small DIG values signal chaotic behavior.

We compute the DIG on a dense grid of initial conditions in $(r,p_r)$ phase space to produce two-dimensional color maps of the global dynamics, revealing regular islands, chaotic layers, and their interfaces. This WBA-based DIC method provides a fast, numerically stable alternative to Lyapunov-based diagnostics, well-suited for high-resolution scans of BH phase spaces. For the visualizations, we use a Python implementation inspired by Ref.~\cite{rolim2025pynamicalsys}. Null geodesics are evolved up to $T=200$ with $N=4000$ sampling points per trajectory. The phase space is sampled on a uniform lattice with $N_r=120$ radial and $N_{p_r}=120$ momentum points. For each initial condition $(r,p_r)$, we integrate the orbit, compute the WBAs, and obtain the DIG from differences between successive averages. This yields global DIG maps that sharply highlight transitions between ordered and chaotic regimes as system parameters vary. {It is important to note that the discrete framework employed in this work arises from the Poincar\'e return map and the associated Weighted Birkhoff Averages, which are constructed by sampling the underlying continuous null geodesic flow. No artificial jump discontinuities are introduced in the dynamics. Therefore, the discrete chaotic indicators used here reflect intrinsic properties of the smooth Hamiltonian system rather than engineered discontinuities, as often encountered in hardware-oriented chaotic maps. Consequently, issues related to the physical implementation of such discontinuities in electronic circuits are not directly relevant to the present analysis.}

\begin{figure*}
	\centerline{
		\includegraphics[width=170mm,height=100mm]{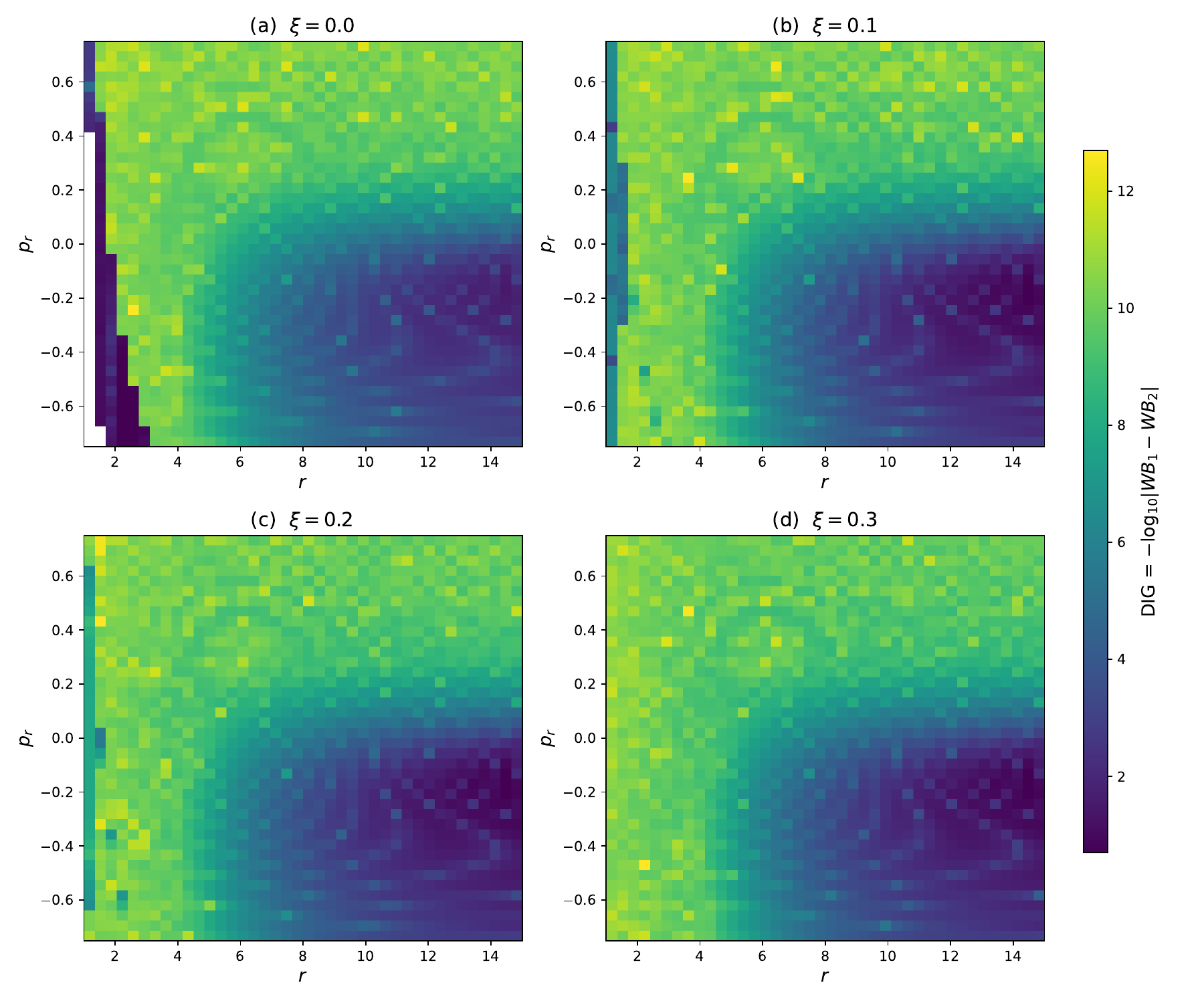}}
	\caption{DIG maps in the $(r,p_r)$ phase space for null geodesics of rotating BHs in WCG for varying $\xi$. Bright (yellow) regions indicate regular motion, while dark (purple) regions correspond to chaotic dynamics. The color bar represents the DIG value $-\log_{10}|WB_1-WB_2|$. The parameters are fixed at $\kappa = 0.01$, $a=0.4$, $\gamma = 0.2$, $E=0.95$ and $L = 3$.}
\label{fig10}
\end{figure*}

Fig.~\ref{fig10} shows the DIG maps in the $(r,p_r)$ phase space for null geodesics of rotating BHs in WCG for different $\xi$, with $\kappa=0.01$, $a=0.4$, $\gamma=0.2$, $E=0.95$, and $L=3$ fixed. All four panels show a smooth, monotonic gradient rather than fragmented islands: bright (high-DIG, regular) regions occupy small-to-intermediate $r$ with positive $p_r$, while dark (low-DIG) regions occupy large $r$ with negative $p_r$, with a gradual transition between them and no visible resonance/tori structure. The only clear $\xi$-dependence is a narrow near-horizon band ($r\lesssim2$): at $\xi=0$ this region shows a pronounced dark strip (and a small unresolved patch), which progressively shrinks as $\xi$ increases to $0.1$, $0.2$, $0.3$, while the rest of the map stays essentially unchanged. This suggests $\xi$ mildly regularizes dynamics near the horizon rather than driving a broad regular-to-chaotic transition. As elsewhere, finite grid resolution and integration time limit how finely narrow chaotic layers within the bright regions could be resolved.

%%%%%%%%%__________________________________SECTION_____________________________________________%%%%
\section{Conclusion and Discussion}\label{S8}
%%%%%______________________________________________________________________________________%%%%
We investigated the dynamical behavior of photon trajectories around a rotating black hole in the framework of Weyl conformal gravity (WCG), focusing on how rotation and conformal corrections modify the phase-space structure and dynamical stability of null geodesics. To characterize the photon dynamics, we employed several complementary diagnostics, including Poincaré sections, finite-time Lyapunov indicators, the Fast Lyapunov Indicator (FLI), finite-time instability measures, and DIG maps. The presence of chaotic and irregular photon motion in the strong-gravity region of black holes is well established in the existing literature and is therefore not claimed here as a new phenomenon. The novelty of the present work lies instead in systematically investigating how the Weyl conformal gravity corrections, encoded by the parameters $\xi$ and $\gamma$, modify the phase-space structure, resonance patterns, and local dynamical instability of photon trajectories in a rotating black hole spacetime. The combined use of complementary diagnostics allows us to examine these WCG-induced modifications from both local and global perspectives and to distinguish the effects of conformal corrections from those associated with black hole rotation.

The Poincaré sections show that photon motion is predominantly regular for relatively small values of the rotation and WCG parameters, with smooth invariant tori associated with quasi-periodic trajectories. As $\xi$, and $\gamma$ increase, these invariant structures become progressively deformed and resonance regions become more pronounced, particularly in the near-extremal rotation regime. The emergence of distorted invariant curves and irregular trajectories indicates an increasing sensitivity to initial conditions, although the results do not imply a complete transition to fully developed global chaos. The WCG parameters modify the effective potential and orbital structure near the photon sphere, with the higher-order curvature parameter $\gamma$ producing stronger deviations from the standard relativistic dynamics and $\xi$ further modifying the phase-space configuration. These effects lead to enhanced nonlinear coupling and increased local dynamical irregularity.

The finite-time Lyapunov indicators and FLI consistently reveal transient variations in the local instability of photon trajectories. Their behavior indicates enhanced sensitivity to initial conditions as the WCG parameters increase, while also emphasizing that these quantities are finite-time diagnostics rather than strict asymptotic measures of chaos. Similarly, the finite-time instability measures exhibit stronger and more persistent fluctuations for larger values of the conformal parameters and near-extremal rotation, supporting the presence of enhanced local instability. The DIG maps provide a complementary global representation, showing a smooth regular-to-irregular gradient across the $(r,p_r)$ phase space rather than fragmentation of regular islands, with the only appreciable $\xi$-dependence being a mild regularization of the dynamics in the narrow near-horizon region rather than a growth of irregular structures toward the photon sphere. {Taken together, these diagnostics indicate that the WCG parameters can significantly modify the local stability and transient dynamics of photon trajectories. However, the decay of the asymptotic Lyapunov exponent toward zero and the non-exponential growth of the FLI do not provide evidence for sustained, fully developed chaotic dynamics in the reduced equatorial system. The observed irregularities are therefore more appropriately interpreted as transient or local instabilities rather than as a definitive transition from regular to chaotic motion.

In summary, our results demonstrate that the dynamical properties of null geodesics in rotating WCG black holes are governed by the interplay between rotation and higher-order conformal corrections encoded in $\xi$ and $\gamma$. Increasing these parameters generally enhances local instability and produces increasingly complex phase-space structures. Such modifications of photon dynamics may have implications for strong-field observables, including photon-ring structures, gravitational lensing, and black-hole shadows, and therefore provide a potential dynamical probe of deviations from general relativity.

The present analysis is restricted to the classical, conservative dynamics of test photons in a prescribed rotating Weyl black-hole spacetime; therefore, Hawking radiation, black-hole thermodynamic entropy, and genuinely nonlocal quantum effects are beyond the scope of this work. Future studies could investigate these effects and examine whether quantum, thermodynamic, or nonlocal corrections modify the dynamical structures identified here. In addition, a more detailed investigation of Hamiltonian phase-space structure, stability, fractal features, and possible chaotic-scattering behavior could provide a deeper understanding of the transition between regular and irregular photon motion.

%\section*{Acknowledgments}
%\textcolor{blue}{The authors would like to thank the anonymous reviewers for their careful reading of the manuscript and for their constructive comments and valuable suggestions, which have significantly improved the quality of this article.}

%\section*{Data Availability Statement}
%The data generated and analyzed in this study are not publicly available due to their size and computational nature. Also, it can be obtained from the corresponding author upon reasonable request.

%\bibliographystyle{apsrev4-1}  %% BibTeX style
%\bibliography{reference}

\bibliographystyle{unsrt}
\bibliography{mainWeyl}

\end{document}